\documentclass[10pt,twoside]{article}
\usepackage{amsfonts,euscript,amssymb,graphicx}
\def\ZZ{{\mathbb Z}}
\newcommand{\be}[1]{\begin{equation}\label{#1}}
\newcommand{\ee}{\end{equation}}
\newcommand{\ba}[1]{\begin{eqnarray}\label{#1}}
\newcommand{\ea}{\end{eqnarray}}
\newcommand{\rf}[1]{(\ref{#1})}
\newcommand{\const}{\mbox{\rm const}\,}
\newcommand{\nn}{\nonumber}

\makeindex

\begin{document}

\title{Phenomenology of Brane-World Cosmological Models}

\markboth{U.\,G\"unther and A.\,Zhuk}{Brane-world phenomenology}

\author{U. G\"unther$^{1,3}$\footnote{present address: Research
Center Rossendorf, P.O. Box 510119, D-01314 Dresden, Germany} \
and A. Zhuk$^{2,3}$
\\[5mm]
\it $^1$Potsdam University, Potsdam, Germany\\
\it e-mail: u.guenther@fz-rossendorf.de\\
\it $^2$Odessa University, Odessa, Ukraine\\
\it e-mail: zhuk@paco.net\\
\it $^3$ICTP, Trieste, Italy}

\date{}
\maketitle

\thispagestyle{empty}

\begin{abstract}
\noindent We present a brief review of brane-world models ---
models in which our observable Universe with its standard matter
fields is assumed as localized on a domain wall (three-brane) in a
higher dimensional surrounding (bulk) spacetime. Models of this
type arise naturally in M-theory and have been intensively studied
during the last years. We pay particular attention to the
covariant projection approach, the Cardassian scenario, to induced
gravity models, self-tuning models and the Ekpyrotic scenario. A
brief  discussion is given of their basic properties and their
connection with conventional FRW cosmology.
\\

\end{abstract}

\section{Introduction \label{intro}}


In the brane-world picture\index{brane world}, it is assumed that
ordinary matter (consisting of the fields and particles of the
Standard Model) is trapped to a three-dimensional submanifold  (a
three-brane) in a higher dimensional (bulk) spacetime\index{bulk
spacetime}. This world-brane is identified with the currently
observable Universe. Such kind of scenarios are inspired by string
theory/M-theory. One of the first examples of this type was the
Ho$\check{\mbox{r}}$ava-Witten setup \cite{HW1} of
$11-$dimensional M-theory compactified on an $S_1/Z_2$
orbifold\index{orbifold} (i.e. a circle folded on itself across a
diameter). In this setup, $E_8$ gauge fields are confined to two
$(1+9)-$dimensional planes/branes located at the two fixed points
of the $S_1/Z_2$ orbifold. In the low-energy limit, the $E_8$
symmetry is down-broken to the $SU(3)\times SU(2)\times U(1)$
group of the Standard Model. Simultaneously, the $9-$brane is
endowed with the structure of a product manifold $M_3\times M_6$
consisting of a noncompact real 3D manifold $M_3$ and a compact
real 6D manifold $M_6$ with the complex structure of a Calabi-Yau
threefold $M_6\sim CY_3$ and  a characteristic compactification
scale $r_{CY_3}$. When the distance between the planes is much
larger than the compactification scale, $R
>> r_{CY_3}$,  then the $CY_3$ degrees of freedom can be integrated out
(see e.g. \cite{Lukas1}) and one arrives at an effective 5-D
heterotic M-theory with two $(1+3)-$dimensional planes.
\begin{figure}[htb]
\centerline{\hbox{\includegraphics[width=0.5\textwidth]{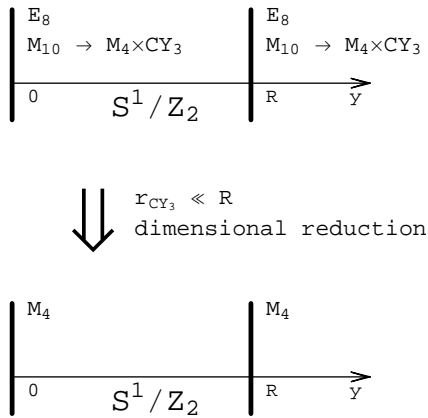}}}
\caption{Schematic of the dimensional reduction of a heterotic
$S^1/\ZZ_2$ orbifold setup.}
\end{figure}
One of these planes describes our brane/Universe and the other one
corresponds to a hidden brane (see Fig. 1). In the formal limit
$R\to \infty$, one obtains a model with a single brane.

The described scenario is one of the possible scenarios for an
embedding of branes into a higher-dimensional bulk manifold.
Because up to now only certain aspects of the underlying
fundamental higher dimensional theory are understood, the model
building process is still going on and an uncovering of new
aspects of brane nucleation/embedding mechanisms can be expected
also for the future.

The presently known brane world models can be distinguished by the
number of branes involved, their co-dimensions, as well as the
topological structure of the branes and the higher dimensional
bulk spacetime. One of the basic tests that any brane-world model
has to pass, is the test on its compatibility with observable
cosmological data and their implications. From the large number of
currently present scenarios we select in the present mini-review
only a few ones --- to illustrate some of the possible deviations
from usual FRW cosmology. The aim of the brief discussion is
twofold. Firstly, for each of the selected models we briefly
review the modifications of the Friedmann-like cosmological
equations and demonstrate the regimes where the corresponding
physics coincides with conventional FRW cosmology; and, secondly,
we briefly indicate the type of observable deviations from it.


\section{4D projected equations along the brane \label{projected}}

Let us consider a single $3-$brane embedded in a five-dimensional
bulk spacetime. Denoting the Gaussian normal (fifth) coordinate
orthogonal to the brane by $y$ we assume  the brane position at
some fixed position $y=\const$ which without loss of generality
can be set as $y=0$. Furthermore, we assume that there is bulk
matter present in 5D and 4D matter on the brane. $\Lambda_5$
denotes the bulk cosmological constant and $\lambda$ the
cosmological constant on the brane (the brane tension). The 5D
Einstein equations for this system read
\be{1.1}
{}^{(5)}G_{AB} = -\Lambda_5 {}^{(5)}g_{AB} +\kappa^2_5
\left[{}^{(5)}T_{AB}+ \left(T_{AB} - \lambda \, g_{AB}
\right)\delta (y)\right]\, .
\ee
The components along the brane can be easily derived by projecting
out the normal components. With the ansatz  $n_AdX^A = dy $, where
$n^A$ is the unit normal to the surface (brane) at $y=0$, the
metric can be split as
\be{1.2}
{}^{(5)}g_{AB} = g_{AB} + n_An_B \; \Longrightarrow \; {}^{(5)}g =
g_{\mu \nu}(x^{\alpha}, y)\; dx^{\mu}\otimes dx^{\nu} + dy \otimes
dy\, .
\ee
The Gauss-Codazzi equations yield then the effective 4D Einstein
equations along the brane (for the details we refer to \cite{SMS}
and the reviews \cite{Maart1,Maart2})
\ba{1.3}
G_{\mu \nu} &=& -\frac12 \Lambda_5 g_{\mu \nu} +\frac23 \kappa^2_5
\mathcal{F}_{\mu\nu}({}^{(5)}T) + K K_{\mu \nu} -
K_{\mu}{}^{\alpha}K_{\alpha \nu} \nn\\
&& +\frac12 \left[K^{\alpha
\beta}K_{\alpha\beta} -K^2\right]g_{\mu \nu} - \mathcal{E}_{\mu
\nu}\, ,
\ea
where
\be{1.4}
\mathcal{F}_{\mu\nu}({}^{(5)}T) \equiv {}^{(5)}T_{AB}g_{\mu}{}^A
g_{\nu}{}^B + \left[ {}^{(5)}T_{AB}n^An^B -\frac14
{}^{(5)}T\right]g_{\mu \nu}
\ee
is the contribution of the bulk matter,
\be{1.5}
\mathcal{E}_{\mu \nu} \equiv
{}^{(5)}C_{ACBD}n^Cn^Dg_{\mu}{}^Ag_{\nu}{}^B
\ee
--- the projection of the bulk Weyl tensor (orthogonal to $n^A$) and
\be{1.6}
K_{AB} = g_A{}^C{}^{(5)}\nabla_Cn_B
\ee
--- the extrinsic curvature of the surface $y=0$.

The on-brane matter can be taken into account via the Israel
junction condition
\be{1.7}
\left. K_{\mu\nu}\right|_{brane} =
-\frac12\kappa^2_5\left[T_{\mu\nu}+\frac13(\lambda
-T)g_{\mu\nu})\right]\, .
\ee
Substitution of this condition into Eq. \rf{1.3} yields the
induced field equation on the brane:
\be{1.8}
 G_{\mu \nu} = -\Lambda g_{\mu\nu} +\kappa^2_4 T_{\mu\nu} +
 \underline{\kappa^4_5 S_{\mu\nu}(T^2)
 - \mathcal{E}_{\mu\nu}({}^{(5)}C) +
\frac 23 \kappa^2_5\mathcal{F}_{\mu\nu}({}^{(5)}T)}\,
 .
\ee
The underlined terms are the specific  brane-world contributions
which are not present in usual general relativity. The first of
these terms,
\be{1.9}
S_{\mu\nu}(T^2) \equiv \frac{1}{12}T T_{\mu\nu}-\frac14
T_{\mu\alpha}T^{\alpha}{}_{\nu}
+\frac{1}{24}g_{\mu\nu}\left[3T_{\alpha\beta}T^{\alpha\beta}-T^2\right]\,
,
\ee
describes local bulk effects (local, because it depends only on
the matter on the brane). The corresponding quadratic corrections
follow from the quadratic extrinsic curvature terms in Eq.
\rf{1.3} and the Israel junction condition \rf{1.7}. The second
and third underlined terms represent non-local bulk effects, with
the traceless contribution $\mathcal{E}_{\mu \nu}$ playing the
role of a {\it dark radiation}\index{dark radiation} on the brane.
The 4D effective cosmological and gravitational constants are
defined as
\be{1.10}
\Lambda := \frac12 \left[\Lambda_5 +\kappa^2_4\lambda\right] \quad
\mbox{and} \quad \kappa^2_4 := \frac16 \lambda \kappa^4_5\, .
\ee
From the latter relation we are lead (for this concrete type of
models) to the following three conclusions. Firstly, only on
positive-tension branes $(\lambda
>0)$ gravity has the standard sign $\kappa^2_4
>0$. A negative tension brane with
$\lambda <0$ would correspond to an anti-gravity world. Secondly,
the on-brane matter gives a linear contribution to the generalized
gravity equations \rf{1.8} only in the case of a non-vanishing
brane tension. A vanishing brane tension $\lambda=0$ implies via
vanishing effective gravitational constant $\kappa_4 ^2=0$ a
vanishing linear energy-momentum contribution, $\kappa^2_4T_{\mu
\nu}=0$, so that only quadratic  terms will remain --- what would
spoil a transition to conventional cosmology in this limit.
Thirdly, Eq. \rf{1.8} reproduces standard Einstein gravity in the
simultaneous limit $\kappa^2_5 \to 0$, $\lambda \to \infty$ for
$\kappa^2_4$ kept fixed and $\mathcal{E}_{\mu\nu}({}^{(5)}C)\to
0$.

Furthermore, it can be shown that for absent bulk matter
(${}^{(5)}T_{AB} = 0$) there will be no leakage of usual matter
from the brane into the bulk,
\be{1.11}
{}^{(4)}\nabla^{\nu}T_{\mu\nu}=0\, ,
\ee
and that in this case the on-brane Bianchi identities set the
following constraint on the dark radiation:
\be{1.12}
{}^{(4)}\nabla^{\mu}\mathcal{E}_{\mu\nu} =
\frac{6\kappa^2_4}{\lambda}{}^{(4)}\nabla^{\mu}S_{\mu\nu}\, .
\ee

Cosmological implications of the model are most easily studied
with the help of a  perfect fluid ansatz for the on-brane energy
momentum tensor (EMT)
\be{1.13}
T_{\mu\nu} = (\rho +P)u_{\mu}u_{\nu} +Pg_{\mu\nu}\, .
\ee
Assuming, furthermore, a natural equation of state for the dark
radiation, $P^{\ast} = (1/3)\rho^{\ast}$, the corresponding energy
momentum contribution can be approximated as
\be{1.14}
-\frac{1}{\kappa^2_4}\mathcal{E}_{\mu\nu} = \frac13
\rho^{\ast}(4u_{\mu}u_{\nu}+g_{\mu\nu})\, .
\ee

Substitution of Eqs. \rf{1.13} and \rf{1.14} into Eq. \rf{1.8}
shows that it is convenient to introduce an effective perfect
fluid with total energy density and pressure
\be{1.15a}
\rho_{tot} :=
\rho\left(1+\frac{\rho}{2\lambda}+\frac{\rho^{\ast}}{\rho}\right)\,
,\qquad  P_{tot} := P + \frac{\rho}{2\lambda}\left(2P+\rho\right)
+\frac{\rho^{\ast}}{3}
\ee
and an effective equation of state
\be{1.17}
\omega_{tot} = \frac{P_{tot}}{\rho_{tot}} =\frac{\omega +
(1+2\omega)\rho/2\lambda +\rho^{\ast}/3\rho}{1 + \rho/2\lambda
+\rho^{\ast}/\rho}\, ,
\ee
where $\omega = P/\rho$. The second and the third terms in Eqs.
\rf{1.15a} 
correspond to local and non-local bulk corrections. These
corrections will be small (and Eq. \rf{1.17} will tend to the
ordinary equation of state $\omega_{tot} \approx \omega$) in the
limit of a large brane tension $\lambda$ and a small energy
density $\rho^{\ast}$ of the dark radiation: $\rho^{\ast} \ll \rho
\ll \lambda$. One expects this to happen, e.g., during late-time
stages of the evolution of the Universe. The considered
brane-world model predicts strong deviations  from conventional
cosmology in the high-energy limit. Assuming that the
phenomenological description via Eqs. \rf{1.8}, \rf{1.15a},
\rf{1.17} is still applicable to energy densities $\rho
>> \lambda , \rho^{\ast}$ at early evolution stages of the
Universe, one obtains equation of state parameters $\omega_{tot}
\approx 2\omega +1$ (i.e. $\omega_{tot} = 5/3$ for $\omega = 1/3$
and $\omega_{tot} = 1$ for $\omega =0$) which would imply
inflation regimes quite different from those in usual 4D FRW
models. We will demonstrate this fact explicitly below.

First, endowing the brane with a FRW metric, we express the
conservation equations \rf{1.11} and \rf{1.12} in terms of the
energy densities:
\be{1.18}
\dot \rho + 3H(\rho+P) = 0 \quad \Longrightarrow \quad \rho \sim
\frac{1}{a^{3(\omega +1)}}
\ee
and
\be{1.19}
\dot \rho^{\ast} + 4H\rho^{\ast} = 0 \quad \Longrightarrow \quad
\rho^{\ast} \sim \frac{1}{a^4}\, ,
\ee
where, as usual, the Hubble constant is given as $H = \dot a /a$,
and overdots denote derivatives with respect to the
synchronous/cosmic time on the brane.

Subsequently, we restrict our attention to the simplest case of a
model without bulk matter ${}^{(5)}T_{AB}=0$ and dark radiation
$\mathcal{E}_{\mu\nu}=0 \Rightarrow \rho^{\ast} =0$ (a model with
vanishing projection of the bulk Weyl tensor onto the brane). Then
the generalized Friedmann equation is of the type
\be{1.20}
H^2 = \frac13
\kappa^2_4\rho\left(1+\underline{\frac{\rho}{2\lambda}}\; \right)
-\frac16{}^{(3)}R +\frac13\Lambda\, .
\ee
As above, we underlined the specific local bulk-induced
contribution which is responsible for the deviation from the
standard Friedmann equation\footnote{Here, the standard Friedmann
equation follows from general relativity with $\Lambda$-term.}.
Again, we observe that this deviation becomes dominant in the
high-energy limit $\rho \gg \lambda$, whereas conventional
cosmology is reproduced at late times when $\rho \ll  \lambda$.

In the simple case of a flat brane ${}^{(3)}R=0$ with vanishing
effective cosmological constant $\Lambda =0$ (compensation of
brane tension $\lambda$ and on-brane effects of the bulk
cosmological constant $\Lambda_5$ according to Eq. \rf{1.10}) the
solution of Eq. \rf{1.20} is easily found explicitly (see
\cite{CS}) as:
\be{1.21}
a(t) = (t-t_0)^{1/[3(\omega +1)]}(t+t_0)^{1/[3(\omega +1)]} \, ,
\quad t_0 := \frac{1}{3(\omega
+1)}\left(\frac{\kappa_5}{\kappa_4}\right)^2\, .
\ee
At early times $t\to t_0$ the scale factor behaves as $a\sim
(t-t_0)^{1/[3(\omega +1)]}$, whereas for late times $t\gg t_0$ the
conventional FRW behavior is restored: $a\sim t^{2/[3(\omega
+1)]}$. Here, the parameters of the model should be chosen in such
a way that the transition to conventional cosmology occurs before
BBN.

As next, we briefly analyze the influence of the additional
contributions of local bulk effects in Eqs. \rf{1.8} and \rf{1.20}
on inflation. We demonstrate the basic effect with the help of the
same simple flat-brane model as above $({}^{(3)}R=\Lambda =0)$ ---
assuming that the energy density and pressure in the state
equation are those of a homogeneous minimally coupled scalar field
which lives on the brane: $\rho = (1/2)\dot \phi^2 + V(\phi)$, $P
= (1/2)\dot \phi^2 - V(\phi)$. The conservation equation \rf{1.11}
${}^{(4)}\nabla^{\nu}T_{\mu\nu}=0 \Rightarrow \dot \rho +
3H(\rho+P) = 0 $ is then equivalent to the field equation $\ddot
\phi + 3H\dot \phi + V' =0$. For the Friedmann equation one gets
\ba{1.22} H^2 &=&
\left[\frac{8\pi}{3M^2_{Pl}}\right]\frac{\rho^2}{2\lambda} \quad
\Rightarrow \quad H(1-\frac13\epsilon ) =
\sqrt{\frac{4\pi}{3M^2_{Pl}\lambda}}\, V(\phi )\equiv A V(\phi
)\, ,\\
\epsilon &\equiv &
\frac{1}{6A}\frac{1}{H}\left(\frac{H'}{H}\right)^2 \approx
\frac{2\lambda}{V}\frac{M^2_{Pl}}{16\pi}\left(\frac{V'}{V}\right)^2\,\nn
\ea
in the high energy limit $\rho \gg \lambda$ and
\ba{1.23}
H^2 &=& \left[\frac{8\pi}{3M^2_{Pl}}\right]\rho \quad \Rightarrow
\quad H^2(1-\frac13\epsilon_c ) = \frac{8\pi}{3M^2_{Pl}}\, V(\phi
)\,
,\\
\epsilon_c &\equiv &
\frac{M^2_{Pl}}{4\pi}\left(\frac{H'}{H}\right)^2 \approx
\frac{M^2_{Pl}}{16\pi}\left(\frac{V'}{V}\right)^2\, \nn
\ea
in the limit of a conventional FRW setup (i.e. in the formal limit
$\lambda \to \infty$). In these equations, the gravitational
constant has been expressed in terms of the Planck mass,
$\kappa^2_4 = 8\pi /M^2_{Pl} = 8\pi G_N$, with  $G_N$ denoting the
Newton constant. The important point is the relation between the
slow-roll parameters $\epsilon$ and $\epsilon_c$ corresponding to
these two limiting cases
\be{1.24}
\epsilon \approx \frac{2\lambda}{V(\phi )}\epsilon_c \, .
\ee
During inflation the high-energy limit with $\rho \approx V(\phi )
\gg \lambda$ holds so that $\epsilon \ll \epsilon_c$ and the local
bulk corrections lead to a strong relaxation of the slow-roll
conditions \cite{HL}. Additionally we note that the deviation of
the cosmological equation \rf{1.22} from the conventional one
\rf{1.23} will necessarily lead to a measurable imprint in the CMB
anisotropies (corresponding details are discussed, e.g., in Ref.
\cite{Maart2}).


\section{Induced gravity\index{induced gravity} brane-world models\label{induced}}

In the previous section, the 4D on-brane gravitational equations
were obtained via the projection of the 5D curvature along the
brane. Another approach is based on the assumption that 4D
effective gravity for matter fields confined to the brane can be
induced on the brane by interactions with 5D gravity in the bulk
[8--15]. 
This results in an effective 4D scalar curvature term ${}^{(4)}R$
(as well as higher order curvature corrections) in the effective
on-brane action functional. For models with vanishing bulk
cosmological constant $(\Lambda_5=0)$ and brane tension
$(\lambda=0)$ it was shown by Dvali, Gabadadze and Porrati (DGP)
\cite{DGP} how quantum interactions can induce effective 4D
on-brane gravity. In their setup, the values of the 5D and 4D
gravitational constants do not dependent on each other, rather
they define a characteristic length scale at which the
gravitational attraction experienced by on-brane matter changes
from 4D Newton's law to that in 5D (see below). Cosmological
implications of induced gravity models were considered in
[10,12,14,16--19]. 
Subsequently, we describe specific issues of induced gravity
cosmology along the lines of Refs. \cite{Sh2,SS}.

Let us consider a model with the following action:
\ba{2.1} S &=&
\frac{1}{2\kappa^2_5}\left[\int_{M_5}d^5x
\sqrt{|{}^{(5)}g|}\left({}^{(5)}R -2\Lambda_5\right)
-2\int_{M_4}d^4x\sqrt{|{}^{(4)}g|}K\right] \nn\\
&&+ \int_{M_5}d^5x \sqrt{|{}^{(5)}g|}L_5 \nn \\ &&+
\frac{1}{2\kappa^2_4}\int_{M_4}d^4x\sqrt{|{}^{(4)}g|}{}^{(4)}R +
\int_{M_4}d^4x \sqrt{|{}^{(4)}g|}\left(L_4 -\lambda\right) \, ,
\ea
where $M_4$ denotes the 4D brane\footnote{In this section, the
brane is assumed as a boundary of a $M_5$ manifold. The theory can
be easily extended to the case where $M_4$ is embedded into
$M_5$.}, and $n_a$ is the vector field of the inner normal to the
brane. The gravitational constants $\kappa^2_{5,4}$ are given in
energy units as: $\kappa^2_{5} = 8\pi /M^3_{\ast}$ and
$\kappa^2_{4} = 8\pi /M^2_{Pl}$. Variation of the action \rf{2.1}
with respect to the 5D bulk metric and the 4D induced metric
yields:
\be{2.2}
{}^{(5)}G_{AB} +\Lambda_5 {}^{(5)}g_{AB} =
\frac{8\pi}{M^3_{\ast}}{}^{(5)}T_{AB}\,
\ee
and
\be{2.3} {}^{(4)}G_{AB} =
\frac{8\pi}{M^2_{Pl}}\left({}^{(4)}T_{AB} -\lambda g_{AB}\right) +
\underline{\frac{M^3_{\ast}}{M^2_{Pl}}S_{AB}}\, ,
\ee
where $g_{AB}$ denotes the induced metric \rf{1.2} on the brane
and
\be{2.4}
S_{AB} := K_{AB}-g_{AB}K\, .
\ee
Eq. \rf{2.3} describes gravity on the brane and the underlined
term defines the bulk correction which is new compared to general
relativity. Let us roughly estimate the length scale at which this
term becomes important. Clearly, it can be dropped  when
$(M^3_{\ast}S_{AB})/(M^2_{Pl} {}^{(4)}G_{AB})\ll 1$. Introducing
the characteristic scales $r_{1,2}$ for ${}^{(4)}G_{AB},S_{AB}$,
i.e. ${}^{(4)}G_{AB} \sim r^{-2}_1\, ,\; S_{AB}\sim r^{-1}_2$, and
assuming as usual for cosmological applications $r_1 \sim r_2$, we
find for this ratio $(M^3_{\ast}r^2_1)/(M^2_{Pl}r_2) << 1
\Longrightarrow r << M^2_{Pl}/M^3_{\ast}$. This means that the
correction plays an essential role at length scales $r \gtrsim l
:= 2M^2_{Pl}/M^3_{\ast}$. Explicitly, one has the following
estimates:
\be{2.4a}
l\sim\left\{
\begin{array}{l}
  10^{15} \mbox{cm} \sim 10^{-3}\mbox{pc}\\
  10^{24} \mbox{cm}\sim 1\mbox{Mpc}\\
  10^{60}\mbox{cm} \gg H_0^{-1}
\end{array}\right. \quad \mbox{for} \qquad M_{\ast} \sim\begin{array}{r}
  1\mbox{TeV}\, , \\
  1\mbox{GeV}\, , \\
  10^{-3}\mbox{eV}\, .
\end{array}
\ee
For energies  $M_{\ast} \sim 10^{-4/3}$GeV the characteristic
length scale is of the order of the present horizon scale $l\sim
H_0^{-1}$, for higher energies it is smaller, and for the dark
energy scale  $M_{\ast} \sim 10^{-3}$eV it is much larger than
$H_0^{-1}$. Hence, in the latter case and, in general, for
$M_{\ast} < 10^{-4/3}$GeV the corrections will be unobservable at
present time. The crucial role of the distance $l$ can be
demonstrated on the behavior of the Newtonian
potential\index{Newton potential}. In Refs. \cite{DGP,DDG}, it was
approximated as
\be{2.5}
V(r) \approx -\frac{G_N}{r}\frac{2}{\pi}\left\{ \frac{\pi}{2} +
\left[ -1+\gamma +\ln\frac{4r}{l}\right]\frac{4r}{l} +
\mathcal{O}(r^2)\right\}\, , \quad r \ll l\, ,
\ee
(effective 4D Newtonian scaling) and as
\be{2.6}
V(r) \approx -\frac{G_N}{r}\frac{2}{\pi}\left\{\frac{l}{4r}+
\mathcal{O}(\frac{1}{r^2})\right\}\, , \quad r \gg l\, ,
\ee
i.e. at large distances $r \gg l$ the gravitational potential
scales as $1/r^2$ in accordance with the laws of a 5D theory.

Let us now illustrate the cosmological implication of this
scenario. For this purpose we consider the simplest model --- a
setup without bulk matter: ${}^{(5)}T_{AB}=0$. The Codazzi
relation guarantees then the conservation of matter on the brane:
\be{2.7}
{}^{(4)}\nabla^{\mu} T_{\mu\nu} = 0\, .
\ee
Here, ${}^{(4)}\nabla$ is the covariant derivative associated with
the induced metric ${}^{(4)}g_{\mu\nu} \equiv g_{\mu\nu}$ (in
normal Gaussian coordinates \rf{1.2}) and we use the notations
${}^{(4)}T_{\mu\nu}\equiv T_{\mu\nu}$, ${}^{(4)}G_{\mu\nu}\equiv
G_{\mu\nu}$. The complete trace of the Gauss relation yields
\be{2.8}
{}^{(4)}R -2\Lambda_5 + S_{AB}S^{AB}-\frac13 S^2 = 0\, ,
\ee
and plugging $S_{AB}$ from Eq. \rf{2.3} into this relation gives
the following on-brane equations system:
\ba{2.9}
&{}& \left[ M^2_{Pl}G_{\mu\nu} + 8\pi \left(\lambda
g_{\mu\nu}-T_{\mu\nu}\right)\right]\left[ M^2_{Pl}G^{\mu\nu} +8\pi
\left(\lambda g^{\mu\nu}-T^{\mu\nu}\right)\right]\nn \\&{}& =
\frac13\left[ M^2_{Pl}{}^{(4)}R - 8\pi\left( 4\lambda -
T\right)\right]^2 -M^6_{\ast}\left( {}^{(4)}R-2\Lambda_5\right)\,
.
\ea
Although these equations seem to form a closed system (they
contain only on-brane quantities), the higher order derivatives
allow for additional degrees of freedom which should be fixed via
integration. One such term which reappears in the corresponding
cosmological equations is the dark radiation. With the help of a
perfect fluid ansatz (similar to that in the previous section), it
was shown in Ref. \cite{Sh2} that integration of Eq. \rf{2.9}
yields:
\be{2.10}
\frac{1}{\kappa^4_4}\left[H^2 +\frac{k}{a^2}
-\frac{1}{3}\kappa^2_4\left(\rho +\lambda\right) \right]^2 =
\frac{1}{\kappa^4_5}\left(H^2 + \frac{k}{a^2}
-\frac{\Lambda_5}{6}-\frac{\mathcal{C}}{a^4}\right)\, ,
\ee
where $k =\pm 1, 0$ and $\mathcal{C}$ is a constant of
integration\footnote{It was shown in Ref. \cite{Sh2}, that the
generalized on-brane Friedmann  equation \rf{1.20} (obtained by
the projection method of the previous section \ref{projected}) can
be recovered from Eq. \rf{2.10} by the formal limit $\kappa_4^2
\to \infty$. For a model with $\ZZ_2-$symmetry and a brane
embedded in the bulk (and not located on the bulk boundary as
considered in the present section) this formal $\kappa_4^2\to
\infty$ limit gives $$H^2 +\frac{k}{a^2} =
\frac{1}{36}\kappa^4_5\left(\rho +\lambda \right)^2 +
\frac16\Lambda_5 + \frac{\mathcal{C}}{a^4} $$ what via relations
\rf{1.10} exactly reproduces \rf{1.20} with additional dark
radiation term.}. Obviously, the term $\mathcal{C}/a^4$ can be
identified as dark radiation. Assuming as before that the 5D and
4D gravitational constants $\kappa^2_{5} = 8\pi /M^3_{\ast}$ and
$\kappa^2_{4} = 8\pi /M^2_{Pl}$ can be fixed independently, the
solution of Eq. \rf{2.10} can be resolved for the Hubble
parameter,
\be{2.14}
H^2 + \frac{k}{a^2} = \frac16\Lambda_5 + \frac{\mathcal{C}}{a^4} +
\frac{1}{l^2}\left[\sqrt{1+l^2\left(\frac13 \kappa^2_4\left(\rho
+\lambda\right)- \frac16\Lambda_5 -
\frac{\mathcal{C}}{a^4}\right)}\; \pm 1 \right]^2\, ,
\ee
where $l=2M^2_{Pl}/M^3_{\ast} = 2\kappa^2_5 /\kappa^2_4$.

Again, the dynamics of on-brane matter depends on the
characteristic length scale $l$. At short distances $r\ll l$, what
corresponds to high energy densities, the bulk corrections play no
important role. Eq. \rf{2.14} shows that in the (UV) limit $\rho
\rightarrow \infty$ we restore the equation of conventional
cosmology
\be{2.15}
H^2 + \frac{k}{a^2} \approx \frac13 \kappa^2_4 \rho\, .
\ee
In the particular case $\lambda =\Lambda_5 = \mathcal{C} =0$ this
happens already at\footnote{This characteristic energy density
changes within the limits $10^{-93}\mbox{g}/\mbox{cm}^3 \leq
\rho_c\leq 10^{-3}\mbox{g}/\mbox{cm}^3$ depending on the concrete
value of $M_{\ast}$ from the interval $10^{-3}\mbox{eV}\leq
M_{\ast}\leq 1\mbox{TeV}$. In the special case  $M_{\ast} \sim
10^{-4/3}$GeV it holds  $\rho_c \sim \rho_0 \sim
10^{-29}\mbox{g}/\mbox{cm}^3$.} $ \rho \gtrsim \rho_c \equiv
1/(\kappa_4 l)^2 \sim M^6_{\ast}/M^2_{Pl}$.

In the opposite case of the IR limit with low energy densities
(i.e. $\rho \rightarrow 0$) the bulk corrections become dominant.
Depending on the sign in Eq. \rf{2.14} two regimes can be
distinguished. (Again, we perform the corresponding sketchy
estimates for the simplest model with  $\lambda =\Lambda_5 =
\mathcal{C} =0$ where the low energy density regime corresponds to
$\rho < \rho_c$.) In the first regime with "minus" sign in Eq.
\rf{2.14} one finds
\be{2.16}
H^2 + \frac{k}{a^2} \approx \frac19 \kappa^4_5 \rho^2\, ,
\ee
which formally coincides with the high energy limit $\rho \gg
\lambda$ in Eq. \rf{1.20} of the previous section.  The different
numerical prefactors 1/9 in Eq. \rf{2.16} and 1/36 in Eq.
\rf{1.20} are the result of different setups: in the previous
section the brane was embedded into the bulk, whereas here it is a
boundary of the bulk. For a flat brane,  $k=0$, the scale factor
behaves non-conventionally: $a\sim t^{1/[3(\omega +1)]}$. One has
to conclude that a Universe (brane-world) which is filled with
ordinary matter (with $\omega >0$) undergoes no late time
acceleration in this regime. The situation is completely different
for models described by Eq. \rf{2.14} with  "plus" sign. Here, the
low energy density regime $\rho << \rho_c$ is necessarily
connected with a dynamics  of the type
\be{2.17}
H^2 + \frac{k}{a^2} \approx \frac{\kappa^4_4}{\kappa^4_5}\, .
\ee
Thus, in this scenario a phase of matter or radiation dominated
cosmology is followed by a late phase of accelerated expansion
\cite{D}.

A more detail analysis of the cosmological behavior in models with
induced gravity can be found, e.g., in Refs.
\cite{DDG,SS}\footnote{It was shown in \cite{SS} that conventional
on-brane dark matter can lead to a cosmology with equation of
state parameter $\omega <-1$ but without phantom-like future
singularity\index{future singularity} as well as with a transient
acceleration with subsequent matter dominated regime.
Additionally, it should be noted that in spite of the appealing
features of the DGP model, it predicts \cite{LPR,R} intensive and
strong interactions at the energy scale $(M_{Pl}/l^2)^{1/3}$,
which for $l\sim H_0^{-1}$ corresponds to distances $\sim
1000\,$km \cite{LPR}. Obviously, this would conflict with
experimental data. However, it was noted in Ref. \cite{NR} that
this problem can be cured with the help of a proper UV completion
of the theory and it may not occur for certain configurations of
the general type \rf{2.1} (with $\Lambda_5,\lambda \neq 0$)
\cite{LPR}.}. An interesting generalization of  DGP-type models to
branes with co-dimensions $D\geq 2$ was proposed in Ref.
\cite{DGHS}. In the case of a co-dimension 2 model, the
characteristic scale at which gravity becomes six-dimensional is
$l\sim {M_{Pl}/M^2_{\ast}}$, which is of order of the Hubble
radius $H_0^{-1}$ for the dark energy mass scale $M_{\ast} \sim
10^{-3}$eV. However, it was pointed out \cite{DR} that such models
are not free of ghosts and will lead to violations of the
equivalence principle.


\section{Cardassian expansion\label{cardassian}}

The discovery of the present late-time acceleration of the
Universe posed a great puzzle to  modern theoretical cosmology. Up
to now several scenarios have been proposed as possible
resolution, such as a cosmological constant, a decaying vacuum
energy and quintessence. However, non of them is fully
satisfactory. On the other hand, as we have seen above, a possible
explanation could consist in an IR modification of gravity and
with it in a modification of the effective Friedmann equation (see
e.g. the low energy limit of Eq. \rf{2.14}). A natural
phenomenological ansatz for an effective Friedmann equation could
consist, e.g., in adding a nonlinear term in the energy density,
\be{3.1}
H^2 = \frac13 \kappa_4^2 \rho +B \rho^n\, ,
\ee
where $B$ is a constant prefactor.  Assuming the matter as dust
with $P=0\, ,\; \rho \sim a^{-3}$ this would lead for a dominating
nonlinear energy density term to
\be{3.2}
H^2 \approx B\rho^n \quad \Longrightarrow \quad a \sim
t^{2/(3n)}\, .
\ee
Thus, for $n<2/3$ the Universe would undergo an accelerated
expansion. Therefore, for large energy densities the main
contribution would come from the first term in Eq. \rf{3.1} and
one would recover conventional cosmology. In contrast to this, for
small energy densities the second term would dominate so that
starting from a certain evolution stage the Universe would undergo
a late time acceleration. Such a scenario was dubbed "Cardassian
expansion"\index{Cardassian expansion}\footnote{The name
Cardassian refers to a humanoid race in "Star Trek" whose goal was
to take over the Universe, i.e. an accelerated expansion.}
\cite{FL}. The appealing feature of this ansatz  is in the fact
that the acceleration is explained without involving vacuum
energy. The question is how to derive the specific
(phenomenological) nonlinear contributions in the energy density
from a more fundamental theory (such as string theory/M-theory).

One possible way could consist in taking advantage of bulk
corrections in brane-world models. For the special case with $n=2$
this has been demonstrated in the previous two sections. In Ref.
\cite{CF} it was shown that, in general, it is possible for such
models to get a modified Friedmann equation $H^2 \sim \rho^n$ with
any $n$. The main idea consists in defining the bulk moduli $N, a$
and $b$ in the metric
\be{3.3}
ds^2 = - N^2(t,y) dt^2 + a^2(t,y) d\vec{x}^2 + b^2 (t,y) dy^2
\ee
in such a way that the energy density of on-brane dust behaves
(via the Israel junction conditions) in accordance with Eq.
\rf{3.2}. On the other hand, these moduli specify the bulk EMT via
the 5D Einstein equations (see also \cite{BDL}). The main drawback
of the approach consists in a very clumsy and inelegant form of
this bulk EMT \cite{FL,CF}. It is very difficult to guess what
kind of matter could produce such an EMT. Furthermore, the
combination $\rho + \rho^n$ with arbitrary $n$ is not yet obtained
by this method. But precisely this combination is required to
induce a matter dominated cosmology which is followed by a late
phase of accelerated expansion (for $n<2/3$).


\section{Self-tuning Universe\label{tuning}}

One of the most  natural candidates for the dark energy causing
the present late time acceleration of the Universe is a positive
vacuum energy (in other words, a cosmological constant). The
energy density of vacuum fluctuations (with natural cut-off $k_c
\sim M_{Pl}$) can be estimated as
\be{4.1}
<\rho_v> = \int^{k_c}_0 \frac{d^3k}{(2\pi )^3}\frac12
\sqrt{k^2+m^2} \; \sim \; \frac{k^4_c}{16\pi^2 } \; \sim \;
M^4_{Pl}\; = \rho_{Pl} \; \sim \; \left(10^{19}\mbox{GeV}\right)^4
\ee
and is 123 orders of magnitude greater than the observable dark
energy density $\rho \sim (10^{-3}\mbox{eV})^4 \sim
10^{-123}\rho_{Pl}$. Usually, this discrepancy is resolved by an
extra ordinary parameter fine tuning and for this reason it is
well known as the fine tuning problem. However, there seems to
exist a possibility to avoid this problem. The idea again consists
in a modification of the Friedmann equation. Let us suppose that
the standard (flat space) Friedmann equation $H^2=(1/3)8\pi G_N
\rho\; $ is modified in such a way \cite{CM} that it holds
\be{4.2}
H^2 = f(\rho ,P)(\rho + P) + ... \, ,
\ee
where $f(\rho ,P)$ is a smooth function of the energy density
$\rho$ and the pressure $P$, and dots denote subdominant terms of
other type. Then the state equation of the vacuum energy,
$P=-\rho$, shows that its contribution will exactly cancel and
that it will not affect the evolution of the Universe, i.e. its
energy density can be arbitrary large. In general, such a
modification can be in agreement with modern observational data.
It is easily demonstrated on the simplest model with
\be{4.3}
H^2 = 2\pi G_N (\rho + P)\, .
\ee
During the matter dominated stage with $P=0$ this equation reads
$H^2 = 2\pi G_N\rho = (3/4) (1/3) 8\pi G_N \rho$. The additional
factor 3/4 is rather close to 1 and such a deviation from the
standard Friedmann equation is very difficult to extract from
observational data. During the radiation dominated stage with
$P=(1/3)\rho$ the standard equation $H^2=(1/3)8\pi G_N \rho\; $ is
restored, what is very important for the BBN. Scenarios with a
modified Friedmann equation of the type \rf{4.2} were dubbed
"self-tuning Universe"\index{self-tuning Universe}. They were
first proposed in Refs. \cite{ADKS,KSS} for specific brane-world
models. An excellent explanation of the basic mechanisms
underlying this setup was given in Ref. \cite{CM}. In our
subsequent brief outline, we mainly follow this work.

The starting point for the scenarios is a 5D brane-world model
with action
\be{4.4}
S = \int d^5x
\sqrt{|{}^{(5)}g|}\left[\frac{1}{2\kappa^2_5}{}^{(5)}R -\alpha
\left({}^{(5)}\nabla \varphi\right)^2\right] + S_4 \left[\varphi ,
{}^{(4)}g, \Psi_i\right]\, ,
\ee
where $\varphi$ is a dilatonic bulk field and $\Psi_i$ are matter
fields living on the brane. The 5D metric ${}^{(5)}g_{AB}$  in the
Einstein frame is connected with the 5D metric ${}^{(5)}\tilde
g_{AB}$ in  Brans-Dicke (BD) frame as
\be{4.5}
{}^{(5)}\tilde g_{AB} = \left(e^{\beta \varphi}\right)
{}^{(5)}g_{AB}\quad \Longrightarrow \quad {}^{(4)}\tilde
g_{\mu\nu} = \left(e^{\beta \varphi}\right) {}^{(4)}g_{\mu\nu}\, ,
\ee
with the induced metrics ${}^{(4)}\tilde g_{\mu\nu}$ and
${}^{(4)}g_{\mu\nu}$ connected by a similar relation. The setup
allows for two possible types of brane-matter coupling in the
effective 4D action on the brane: either the matter is coupled to
the induced 4D metric in the BD frame,
\be{4.6}
S_{(1)4} = \int d^4x \sqrt{|{}^{(4)}\tilde g|}\; L_4 \left(\Psi_i,
{}^{(4)}\tilde g\right)\, ,
\ee
or it is coupled to the induced metric in the Einstein frame,
\be{4.7}
S_{(2)4} = \int d^4x \sqrt{|{}^{(4)}\tilde g|}\; L_4 \left(\Psi_i,
{}^{(4)}g\right) = \int d^4x \sqrt{|{}^{(4)} g|}\; \left(e^{2\beta
\varphi}\right) L_4 \left(\Psi_i, {}^{(4)} g\right)\, .
\ee
Using the ansatz \rf{3.3} for the 5D metric (in the Einstein
frame) and the special parameter tuning
\be{4.10}
\frac{\beta^2}{\alpha} = \frac23 \kappa^2_5
\ee
one finds the following on-brane  equations
\be{4.8}
\frac{\ddot a_0}{a_0} + \frac{\dot a_0^2}{a_0^2} = \dot H + 2 H^2
= \left\{\begin{array}{rcl}
-\frac{1}{32}\kappa^4_5e^{4\beta\varphi_0} \left(\tilde \rho +
\tilde P\right)^2 &-&\frac12 \beta^2 \dot \varphi^2_0 \, , \quad (\mbox{I})\\
&\phantom{\int}&\\
-\frac{1}{36}\kappa^4_5 \left( \rho + P\right)\left(\rho +
2P\right) &-&\frac12 \beta^2 \dot \varphi^2_0\, ,
\quad (\mbox{II})\\
\end{array}\right.
\ee
which contain energy-density---pressure combinations of a form
similar to the self-tuning combinations of the generalized
Friedmann equation  \rf{4.2}. In \rf{4.8} equation (I) corresponds
to the BD frame coupling and equation (II) to the Einstein frame
coupling. The tilded and non-tilded energy densities and pressures
$(\tilde \rho, \tilde P)$ and $(\rho, P)$ are defined from Eqs.
\rf{4.6} and \rf{4.7} with respect to metrics ${}^{(4)}\tilde
g_{\mu\nu}$ and ${}^{(4)}g_{\mu\nu}$, respectively. 0-subscripts
refer to on-brane quantities.

Comparing equations (I) and (II) in \rf{4.8} with the self-tuning
equations \rf{4.2} and \rf{4.3}, one immediately realizes that the
quadratic energy-density---pressure term in Eq. (I) would yield a
self-tuning but it would spoil the possibility for a limit to
conventional cosmological behavior. In contrast to this, Eq. (II)
can provide the necessary linear dependence. Splitting, for
example, the on-brane energy density and  pressure into a static
background component (which can be identified with a vacuum
contribution\footnote{The self-tuning scenario was designed to
cancel the huge vacuum contributions, but in its current form it
is not yet capable to elegantly induce accelerated expansion ---
like inflation or late-time acceleration (see also footnote
\ref{problems2}).\label{problems1}}) and a dynamical component of
ordinary (non-vacuum) matter,
\be{4.11}
\rho = e^{2\beta \varphi_0}\lambda +\rho_m\, , \quad P = -
e^{2\beta \varphi_0}\lambda + P_m\, ,
\ee
(with $\lambda$ as brane  tension) one gets from Eq. (II) for
constant on-brane values of the dilaton\footnote{Here, the
stabilization problem for the on-brane value of the dilaton is
left aside. It would require a separate investigation.} $(\dot
\varphi_0 = 0)$ an equation capable for a self-tuning
\be{4.12}
\dot H + 2H^2 = \frac{\kappa^4_5}{36}\left[e^{2\beta
\varphi_0}\lambda \left(\rho_m + P_m\right) - \left(\rho^2_m +
3\rho_mP_m + 2P_m^2\right)\right]\, .
\ee
In a late-time regime, the term quadratic in the dynamical energy
density and pressure would be subdominant and it remains to check
whether  the resulting linear model will be capable to reproduce
conventional cosmology. The analysis can be made explicit, e.g.,
by considering the simplest case of a flat-brane ansatz for the
metric \rf{3.3}
\be{4.13}
ds^2 = \omega (y) \left[-dt^2 + a^2_0(t)d\vec{x}^2 \right] +
dy^2\, .
\ee
The concrete form of $\omega (y)$ was found in Refs.
\cite{ADKS,KSS} as
\be{4.14}
\omega (y) = \sqrt{1- y/y_c} \, , \quad y_c =
\frac{3\kappa^{-2}_5}{2\lambda} e^{-2\beta \varphi_0}\, ,
\ee
where $y_c$ corresponds to a boundary (opposite to the
world-brane) at which the metric experiences a singularity,
$\omega (y_c)=0$. The effective 4D gravitational constant on the
brane can be obtained via dimensional reduction as
\be{4.15}
\frac{1}{8\pi G_N} = \frac{1}{\kappa^2_4} = \frac{1}{\kappa^2_5}
\int_0^{y_c} dy\omega (y) = \frac{1}{\lambda\kappa^4_5}
e^{-2\beta\varphi_0}\, .
\ee
Thus, up to quadratic terms, Eq. \rf{4.12} reads
\be{4.16}
\dot H +2H^2 \approx \frac{2\pi G_N}{9}\left(\rho_m + P_m\right)\,
\ee
and yields after integration
\be{4.17}
H^2 = \frac{4\pi G_N}{9}\, a^{-4}\int  a^3 \left(\rho_m + P_m
\right)da\, .
\ee
For a Universe which is dominated by a combination of dust
($\rho_d \sim 1/a^3, P_d = 0$) and radiation ($\rho_r \sim 1/a^4,
P_r = (1/3)\rho_r $) this leads to a Hubble parameter of the type
\be{4.18}
H^2 = \frac13 8\pi G_N \left[\frac16 \rho_d + \frac29 \rho_r
\ln\left(\frac{a}{a_{\ast}}\right)\right]
\ee
with $a_{\ast}$ as constant of integration.

Obviously, the considered toy model does not exactly reproduce the
dynamics of conventional cosmology. During the matter dominated
stage the additional prefactor $1/6$ induces a deviation from
conventional cosmology which probably will leave an imprint in the
CMB anisotropy. Also the dynamics of the radiation dominated stage
is governed by a different equation --- which can affect the BBN.
This may require a special tuning mechanism\footnote{Additionally,
it should be noted that up to now no convincing mechanism has been
found for an elegant matching of the self-tuning scenarios with
inflation. The point is that for conventional cosmology a vacuum
equation of state with $P=-\rho $ yields $H=$const. In contrast to
this, setting in Eq. \rf{4.12} $\rho_m =0= P_m $ (i.e., a vacuum
of the dynamical contributions) gives  $H \sim 1/(2t)$ and $H$
behaves in the same way as during the RD stage of conventional
cosmology. \label{problems2}} to produce the correct light-element
abundances. Nevertheless, the simple toy model allowed us to
illuminate the main ideas of the self-tuning mechanism. Further
details of the scenario have been worked out, e.g., in Refs.
[31--34]. 


\section{Ekpyrotic Universe\label{ekpyrotic}}

A very interesting scenario which is based on a different general
setup --- compared with those in the previous sections --- was
proposed in Ref. \cite{KOST}. It is based on the assumption that
the evolution of our Universe (in accordance with known
observational data) can be explained with the help of a relatively
simple brany M-theory setup with $S^1/Z_2$ orbifold symmetry. The
starting point is the Ho$\check{\mbox{r}}$ava-Witten model
(briefly mentioned in the introduction section \ref{intro}) which
after dimensional reduction leads to an effective 5D action of the
type
\be{5.1}
S = \frac{1}{2\kappa^2_5} \int_{M_5} d^5x \sqrt{|{}^{(5)}g|}
\left[ {}^{(5)}R -\frac12 \left(\partial \varphi\right)^2 -\frac32
\frac{e^{2\varphi}\mathcal{F}^2}{5!}\right] + \mbox{brane terms}\,
.
\ee
The scalar field $\varphi$ defines the volume scale (modulus) of
the Calabi-Yau three-fold and $\mathcal{F}$ is the field strength
of a four-form gauge field. The key point of this model is a
static (nearly) Bogomol'nyi-Prasad-Sommerfeld (BPS) solution
\cite{KOST} which describes a setup consisting of three 3-branes:
\ba{5.2}
ds^2 &=& D(y)\left(-N^2d\tau^2 + A^2d\vec{x}^2\right)
+B^2D^4(y)dy^2\, ,\nn \\
e^{\varphi} &=& B D^3(y) \, ,\\
\mathcal{F}_{0123y} &=& -\alpha A^3NB^{-1}D^{-2}(y)\, , \quad
(y<Y)\,
,\nn \\
&=& -(\alpha-\beta) A^3NB^{-1}D^{-2}(y)\, , \quad (y>Y)\, ,
\ea
where $\tau$ denotes the conformal time, $D(y)$ is defined as
\ba{5.3}
D(y) &=& \alpha y + C\, ,  \quad (y<Y)\, , \\
&=& (\alpha - \beta)y + C +\beta Y\, , \quad (y>Y)\nn
\ea
and $A,B,C,N,Y$ are constants (with $C>0$). Two of the three
branes are located at the orbifold fixed points $y=0$ and $y=R$,
whereas the third (bulk) brane is positioned at $0<y=Y<R$. Our
visible brane (at $y=0$) is assumed to have negative tension
$\alpha_1 \equiv \alpha <0$, the hidden brane (at $y=R$) $\alpha_2
\equiv \alpha -
\beta >0$ and the bulk brane $\alpha_3 \equiv \beta >0$ to have a positive one.
Furthermore, it is assumed that the bulk brane is nucleated near
the positive tension (hidden) brane at $y=R$. It is supposed that
the bulk brane is very light: $\beta << \alpha$. A very important
property of the (nearly) BPS solution is that all these three
branes are flat and parallel one to the other.

\begin{figure}[htb]
\centerline{\hbox{\includegraphics[width=0.5\textwidth]{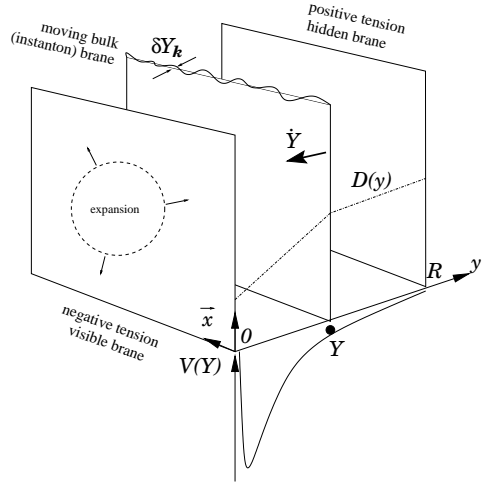}}}
\caption{\label{fig2}Schematic of the ekpyrotic scenario (from
\cite{KKL}).}
\end{figure}

Additionally, it is assumed that a non-perturbative interaction
between the bulk and boundary branes leads to an effective
potential
\be{5.4}
V(Y) \approx  -v e^{-m\alpha Y}\, ,
\ee
which for small $Y$ suddenly becomes zero ($v$ and $m$ are
positive dimensionless parameters of the model). Via the potential
the light bulk brane is attracted to the visible brane and moves
adiabatically slowly towards the visible brane (see the schematic
illustration in Fig. \ref{fig2}).

In this scenario, branes and bulk are assumed to start out from a
cold, symmetric state with \rf{5.2} which is nearly BPS. The
beginning of our Universe, i.e. the Big Bang, is identified in
this model with the impact of the bulk brane on the visible brane.
After this collision, a part of the kinetic energy of the bulk
brane transforms into radiation which is deposited in the three
dimensional space of the visible brane. The scenario was called
"Ekpyrotic Universe"\index{Ekpyrotic Universe}
\cite{KOST}\footnote{This term was drawn from the Stoic model of
cosmic evolution in which the Universe is consumed by fire at
regular intervals and reconstituted out of this fire --- a
conflagration called ekpyrosis. In the present scenario, the
Universe is made through a conflagration ignited by a brane
collision.}. An early detailed analysis of its features
was performed in Refs. [35--41].

It was argued that this model could solve the main cosmological
problems without invoking inflation and as alternative to
it\footnote{See, however, the critical comments in \cite{KKL,KKL2}
and the reply to these comments in \cite{KOST2}.}:
\begin{itemize}
\item It is assumed that the {\it monopole problem} can be avoided if
the parameters of the model are chosen in such a way that the
temperature on the visible brane after collision is not high
enough for a production of primordial monopoles.
\item The {\it flatness and homogeneity problems} are automatically
solved\footnote{For the solution of these problems in other brane
world models, e.g. such with large extra dimensions,  different
smoothing mechanisms have been proposed \cite{SST}.} because the
nearly BPS branes are assumed as sufficiently flat and homogeneous
(see however the critical comments in \cite{KKL,KKL2} concerning
the naturalness of the required high parameter tuning).
\item The {\it horizon problem} is solved by the extremely  slow
motion of the bulk brane so that, before collision, particles on
the visible brane can travel an exponentially long distance. As
result, the horizon distance $d_{hor}$ can be much bigger than the
Hubble radius at collision $H_c^{-1} \sim T_c^{-2}$, where $T_c$
is the temperature of the visible brane at collision.
\item It is assumed that the {\it large-scale structure} on the visible
brane is induced by quantum fluctuations (ripples) in the position
of the bulk brane. These fluctuations would lead to a
space-dependent time delay in the brane collision which on its
turn would result in density fluctuations on the visible brane.
Because small ripples can generate a large time delay, the
collision could induce  a spectrum of density fluctuation which
would extend to exponentially large, super-horizon scales. In the
considered scenario, the spectrum of perturbations is
approximately scale invariant. However, in the simplest variant of
the Ekpyrotic model the spectrum of adiabatic fluctuations is blue
\cite{Lyth,BF}.
\end{itemize}

Summarizing, the M-theory based scenario of an Ekpyrotic Universe
has a large number of interesting features which could present an
alternative to inflation. Future investigations will show which of
the scenarios is more robust. Further developments in this
direction can be found, e.g., in Refs. \cite{KOSST,ST}.


\section{Conclusion\label{conclusion}}

We are living in very interesting time because new observational
data directly indicate that our knowledge about the Universe and
its evolution is in great extent limited. In this situation new
ideas and new theoretical models are required to explain the
observational data. In their turn, these new models will predict
new observable phenomena which can be used as test tools to single
out those scenarios which are most close to nature.  One possible
direction to modify the currently accepted standard theory of
cosmology are brane-world models --- M-theory inspired setups in
which our observable Universe is interpreted as a 3D submanifold
embedded in a higher dimensional bulk spacetime. The aim of the
present mini-review was to give a very brief description of some
of the basic features of these models --- illustrating them with
the help of a few concrete toy model setups.

\vspace*{1ex}

\mbox{} \\ {\bf Acknowledgments}\\ We thank the High Energy,
Cosmology and Astroparticle Physics Section of the ICTP (Trieste)
for their warm hospitality during the preparation of this
mini-review as well as for financial support from the Associate
Scheme (A.Z.) and from HECAP (U.G.). Additionally, U.G.
acknowledges support from DFG grant KON/1806/2004/GU/522.




\end{document}